\documentclass[aps,showpacs]{revtex4}
\usepackage{latexsym, amscd}
\usepackage{amsmath}
\usepackage{graphicx}

\begin{document}

\title{Symmetric and asymmetric solitons trapped in H-shaped potentials}
\author{ Nguyen Viet Hung $^{1,2}$, Marek Trippenbach $^{3}$, and Boris A.
Malomed $^{4,5}$ } \affiliation{$^1$ Soltan Institute for Nuclear
Studies, Ho\.{z}a 69, PL-00-681 Warsaw, Poland \\$^2$ Vinh
University, 182 Le Duan
str., Vinh, Nghe An, Vietnam\\
 $^{3}$ Institute of Theoretical
Physics, Physics Department, Warsaw
University, Ho\.{z}a 69, PL-00-681 Warsaw, Poland\\
$^{4}$ Department of Physical Electronics, School of Electrical
Engineering,
Faculty of Engineering, Tel Aviv University, Tel Aviv 69978, Israel\\
$^{5}$ ICFO-Institut de Ciencies Fotoniques, Mediterranean
Technology Park,
08860 Castelldefels (Barcelona), Spain. \thanks{%
a temporary Sabbatical address}}

\begin{abstract}
We report results of numerical and analytical studies of the spontaneous
symmetry breaking in solitons, both two- and one-dimensional, which are
trapped in H-shaped potential profiles, built of two parallel potential
troughs linked by a narrow rung in the transverse direction. This system can
be implemented in self-attractive Bose-Einstein condensates (BECs), as well
as in a nonlinear bulk optical waveguide.We demonstrate that the
introduction of the transverse link changes the character of the
symmetry-breaking bifurcation (SBB) in the system from subcritical to
supercritical (in terms of the corresponding phase transition, it is a
change between the first and second kinds). A noteworthy feature of the SBB
in this setting is a non-monotonous dependence of the soliton's norm at the
bifurcation point on the strength of the transverse link. In the full 2D
system, the results are obtained in a numerical form. An exact analytical
solution is found for the bifurcation in the 1D version of the model, with
the transverse rung modeled by the local linear coupling between the
parallel troughs with the $\delta $-functional longitudinal profile.
Replacing the $\delta $-function by its finite-width Gaussian counterpart,
similar results are obtained by means of the variational approximation (VA).
The VA is also applied to the 1D system with a mixed linear and nonlinear
transverse localized coupling. Comparison of the results produced by the
different varieties of the system clearly reveals basic features of the
symmetry-breaking transition in it.
\end{abstract}

\maketitle

\section{Introduction}

Symmetric double-well potentials is one of fundamental settings studied in
quantum mechanics \cite{LL} and in the theory of optical guided-wave
propagation, which obeys the Schr\"{o}dinger equation similar to that known
in quantum mechanics \cite{NLS}. Counterparts of the double-well potential
in optics are represented by directional couplers \cite{Snyder,dual-core},
including various dual-core waveguides created in photonic-crystal matrices
\cite{PCFcoupler}. It is commonly known that the ground state produced by
the linear Schr\"{o}dinger equation keeps the symmetry of the underlying
double-well potential \cite{LL}. On the other hand, the introduction of the
self-attractive nonlinearity, which transforms the linear equation into the
Gross-Pitaevskii equation (GPE) for a Bose-Einstein condensate (BEC)\ of
interacting atoms, loaded into the double-well potential \cite{BEC}, or the
nonlinear Schr\"{o}dinger equation (NLSE) modeling nonlinear dual-core
waveguides in optics \cite{NLS}, leads to the ubiquitous effect of the
spontaneous symmetry breaking. As a result, the symmetric ground state is
replaced, via the \textit{symmetry-breaking bifurcation} (SBB; in fact, it
is a variety of phase transitions), by an asymmetric state providing for a
minimum of the system's energy, when the strength of the nonlinearity
exceeds a critical value. This effect was originally discovered in a
discrete model of self-trapping \cite{Chris}, and later studied in many
settings \cite{misc}. Manifestations of the spontaneous symmetry breaking
were also studied in detail in BEC models \cite{misc2} and demonstrated
experimentally in a self-repulsive BEC \cite{Markus}.

In nonlinear optics, the SBB\ was analyzed in Ref. \cite{Snyder} for
continuous-wave (spatially uniform) states in the model of dual-core
waveguides. For self-trapped modes in dual-core optical systems, i.e.,
solitons, the bifurcation was studied in Refs. \cite{dual-core,Amir}. A
specific example is the SBB for gap solitons in dual-core fiber Bragg
gratings \cite{Mak}. Later, manifestations of the SBB\ for matter-wave
solitons held in \textit{dual-trough} potential traps (including solitons of
the gap type, supported by a periodic optical-lattice potential) were
explored too \cite{Arik}-\cite{Luca}. The difference of the dual-trough
configuration from the usual double well is the presence of an additional
free direction, transverse to that in which the double-well potential acts,
see Fig. \ref{potenhshaped} below.

A characteristic property of solitons trapped in the symmetric dual-trough
potential is that, under the action of the self-focusing cubic nonlinearity,
they exhibit the SBB of the \textit{subcritical} type. In that case,
branches of the asymmetric states emerge as unstable ones, going backward
from the SBB\ point and getting stable after switching their direction
forward at turning points \cite{bif}. Thus, the system features the
bistability before the bifurcation point, and this mode of the spontaneous
symmetry change may be understood as the phase transition of the first kind.
On the other hand, the addition of a sufficiently strong periodic potential
acting along the troughs changes the character of the SBB in the
self-attractive medium from subcritical to \textit{supercritical}. In the
latter case, the asymmetric branches emerge as stable ones, immediately
going in the forward direction \cite{Arik,Warsaw2}, which may also be
realized as the phase transition of the second kind. The above-mentioned SBB
for gap solitons in the dual-core fiber Bragg grating is of the
supercritical type too \cite{Mak}.

Another realization of effective double-well potentials is provided by
settings based on double-peak spatial modulations of the local nonlinearity
coefficient, which may be implemented in optics and BEC alike, see recent
review \cite{Barcelona} of the topic of solitons in nonlinear potentials.
The simplest form of this setting has the nonlinearity concentrated at two
points, in the form of a symmetric pair of delta-functions or narrow
Gaussians \cite{Thawatchai}, as well as a two-dimensional (2D) counterpart
of the system, based on the set of two parallel stripes \cite{Hung}, or two
circles \cite{Thaw2D} carrying the self-attractive nonlinearity. The SBB of
solitons in the double-well nonlinear potentials also features the symmetry
breaking of the subcritical type \cite{Thawatchai,Hung,Barcelona}.

Further, the spontaneous symmetry breaking was analyzed for 1D and 2D
solitons in dual-core discrete systems, with the uniform coupling between
two parallel chains \cite{Herring}, or with the coupling concentrated at a
single site \cite{Ljupco}. In the former and latter cases, the SBB is
subcritical and supercritical, respectively. Another implementation of the
SBB in discrete settings was recently reported for a pair of nonlinear sites
embedded into or side-coupled to a linear host lattice \cite{Almas}.

The objective of the present work is to consider the SBB of self-attractive
localized wave fields trapped in H-shaped potential landscapes, i.e., two
parallel troughs linked by a transverse rung, as shown in Fig. \ref%
{potenhshaped} below. This configuration can be implemented in effectively
two-dimensional BEC, using a set of attractive (red-detuned) laser sheets
and/or blue-detuned repelling sheet pairs \cite{sheet}, or in BEC\ layers
isolated by a strong standing optical wave \cite{standing}, that can also be
combined with magnetic trapping fields \cite{2Dcombined}. It is also
possible to use the techniques allowing one to ``paint" complex potential
landscapes (in fact, even more complex than the H-shaped ones that we aim to
consider) by rapidly moving laser beams \cite{writing}, or induce
time-averaged adiabatic landscapes created by means of variable magnetic
fields \cite{TAAP}. Essentially the same effective potentials can be created
in nonlinear optics, using properly patterned photonic-crystal media \cite%
{PCFcoupler}, or a transverse trapping structure permanently written in bulk
silica \cite{Jena}. We aim to consider both the full 2D model with the
transverse H-shaped potential (similar to the 2D model with the dual-trough
potentials considered in Refs. \cite{Warsaw1,Warsaw2}) and its simplified 1D
counterpart (cf. the 1D version of the dual trough introduced in Ref. \cite%
{Arik}).

The transverse link (rung) added to the dual-trough potential can be used to
control the dynamical properties of the system and, eventually, alter the
character of the spontaneous symmetry breaking in the system. In particular,
the recent results reported for the discrete system \cite{Ljupco} suggest a
possibility to change the type of the SBB\ from sub- to supercritical (i.e.,
the kind of the respective phase transitions from first to second) by
gradually increasing the strength of the transverse link.

The paper is structured as follows. The 2D and 1D models are introduced in
Section II, and the full 2D system is considered in Section III by means of
numerical methods. A set of bifurcation diagrams indeed demonstrates a
switch from the sub- to supercritical SBB in the 2D setting. While one may
expect that the strengthening of the linear coupling between the parallel
troughs should lead to an increase of the critical value of the nonlinearity
strength at which the SBB happens, we observe that, with the introduction of
the transverse link, the critical nonlinearity strength at first decreases,
due to the change of the character of the bifurcation, and only later starts
to grow. In Section IV, we deal with the 1D versions of the system. In that
case, the transverse link reduces to a localized linear coupling between two
one-dimensional GPEs/NLSEs. In that context, we consider different
longitudinal profiles of the coupling. First, approximating it by a $\delta $%
-function of the longitudinal coordinate, we report exact analytical
solutions for the solitons of all the types---symmetric, antisymmetric, and
asymmetric. Accordingly, an exact comprehensive solution for the SBB is
available. Next, we consider the coupling localized in a finite interval,
with the Gaussian profile. In that case, we develop a variational
approximation (VA), and verify its predictions by comparison to numerical
findings. Finally, we consider the 1D system which combines the linear and
nonlinear localized couplings between the one-dimensional GPEs/NLSEs, both
with the Gaussian profile. For that purpose, the VA is developed too. In all
the versions of the 1D system, the SBB is always found to be of the
supercritical type, unlike the full 2D model that reveals the transition to
the supercritical type from the subcritical one. The paper is concluded by
Section V.

\section{The model}

\subsection{The two-dimensional setting}

\begin{figure}[tbp]
\begin{center}
\includegraphics[width=8cm]{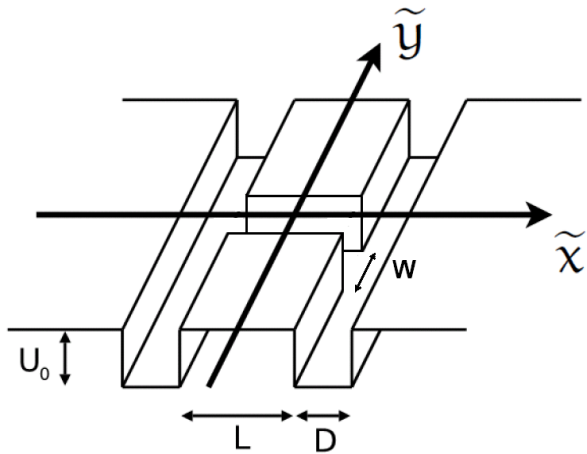}
\end{center}
\caption{The two-dimensional H-shaped trapping potential.}
\label{potenhshaped}
\end{figure}

We introduce the model in terms of the two-dimensional GPE for the
mean-field wave function of the self-attractive BEC trapped in the
H-shaped potential. In physical units, the usual form of the
equation is
\begin{equation}
i\hbar \widetilde{\Psi }_{\widetilde{t}}=-\frac{\hbar ^{2}}{2m}(\widetilde{%
\Psi }_{\widetilde{x}\widetilde{x}}+\widetilde{\Psi }_{\widetilde{y}%
\widetilde{y}})+U(\widetilde{x},\widetilde{y})\widetilde{\Psi }+\frac{2\sqrt{%
2\pi }\hbar ^{2}a_{s}}{ma_{z}}|\widetilde{\Psi }|^{2}\widetilde{\Psi },
\label{2DGPorigin}
\end{equation}%
where $m$ and $a_{s}<0$ are the atomic mass and the respective scattering
length, $a_{z}$ is the confinement length in the transverse direction [$%
a_{z}=\sqrt{\hbar /\left( m\Omega \right) }$, if the confinement is provided
by the harmonic-oscillator potential, $(1/2)m\Omega ^{2}z^{2}$], and the
trapping potential of depth $U_{0}$ is defined as follows:
\begin{equation}
U(\widetilde{x},\widetilde{y})=\left\{
\begin{array}{ll}
-U_{0}, & \mbox{ at }~~\frac{1}{2}L<|\widetilde{x}|<D+\frac{1}{2}L, \\
~~~~~ & ~\mbox{and at}~|\widetilde{x}|<\frac{1}{2}L,~|\widetilde{y}|<\frac{1%
}{2}W, \\
~~~0 & ~\mbox{elsewhere}.%
\end{array}%
\right.   \label{hshaped}
\end{equation}%
As shown in Fig. \ref{potenhshaped}, $D$ is the width of the two
longitudinal troughs, with separation $L$ between them, and $W$ is the width
of the transverse rung. The norm of the wave function gives the total number
of atoms in the condensate,
\begin{equation}
N_{p}=\int \int_{-\infty }^{\infty }|\widetilde{\Psi }|^{2}d\widetilde{x}d%
\widetilde{y}.  \label{2DNormorigin}
\end{equation}%
\newline

Using scaled variables $x=\widetilde{x}/D$, $y=\widetilde{y}/D$, $t=%
\widetilde{t}\hbar /(mD^{2})$, and $\Psi \equiv 2\sqrt{2\sqrt{2\pi }%
\left\vert a_{s}\right\vert /a_{z}}D\widetilde{\Psi }$, we cast Eq. (\ref%
{2DGPorigin}) into the dimensionless form:
\begin{equation}
i\Psi _{t}=-\frac{1}{2}(\Psi _{xx}+\Psi _{yy})+U(x,y)\Psi -|\Psi |^{2}\Psi ,
\label{2DGPtwowell}
\end{equation}%
where the rescaled norm is $N_{\mathrm{2D}}\equiv 4\pi N_{p}|a_{s}|$, and
the potential takes the form of
\begin{equation}
U(x,y)=\left\{
\begin{array}{ll}
-\alpha _{1}, & \mathrm{at~}~\frac{1}{2}\alpha _{2}<|x|<1+\frac{1}{2}\alpha
_{2}, \\
~~~~~ & \mathrm{at~}~|x|<\frac{1}{2}\alpha _{2},~|y|<\frac{1}{2}\alpha _{3},
\\
~~~~~0, & ~\mathrm{elsewhere},%
\end{array}%
\right.   \label{twosquare1}
\end{equation}%
with rescaled constants%
\begin{equation}
\alpha _{1}\equiv \frac{mU_{0}D^{2}}{\hbar ^{2}},~\alpha _{2}\equiv \frac{L}{%
D},~\alpha _{3}\equiv \frac{W}{D}~.  \label{alpha}
\end{equation}%
Thus, the system is governed by the set of four dimensionless parameters:
norm $N_{\mathrm{2D}}$, scaled potential depth $\alpha _{1},$ the relative
separation between the parallel channels, $\alpha _{2}$, and the relative
width of the rung, $\alpha _{3}$. In the next section, we apply a numerical
method to search for symmetric and symmetry-broken localized modes supported
by the interplay of the H-shaped trapping potential and self-attractive
nonlinearity in Eq. (\ref{2DGPtwowell}). We will identify the SBB in this
model, and produce the respective bifurcation diagrams. The predicted
results can be readily translated back into physical units be undoing the
above transformations.

In the application to the optical guided-wave propagation, Eq. (\ref%
{2DGPtwowell}) plays the role of the NLSE for the evolution of the amplitude
of the guided electromagnetic wave, with time $t$ replaced by the
propagation distance, $z$. In the latter case, the effective potential
represents the modulation of the refractive index in the transverse plane,
and the cubic term accounts for the Kerr self-focusing.

\subsection{The one-dimensional system}

The 1D limit of the model corresponds to the case of $W\ll D\ll L$, i.e., $%
\alpha _{3}\ll 1\ll \alpha _{2}$, in terms of the relative parameters
defined in Eq. (\ref{alpha}). Then, approximating the wave function in each
relatively narrow trough by 1D wave functions $\psi _{1,2}\left( y,t\right) $%
, the tails of the wave functions channeled by the narrow transverse rung in
the transverse direction take the form of
\begin{equation}
\left( \psi _{1,2}\right) _{\mathrm{tail}}\left( x,y,t\right) \approx \psi
_{1,2}\left( y,t\right) \exp \left[ -\sqrt{\left( \frac{\pi }{\alpha _{3}}%
\right) ^{2}-2\alpha _{1}}\left\vert x\pm \frac{\alpha _{2}}{2}\right\vert %
\right]  \label{tail}
\end{equation}%
[which implies $\alpha _{1}<\pi ^{2}/\left( 2\alpha _{3}^{2}\right) $;
otherwise (if the rung is very deep), the effective coupling between the
parallel troughs will be stronger than given below by expression (\ref{kappa}%
)]. Then, using the general formalism elaborated for the analysis of the
interaction between far separated 2D localized modes \cite{2D}, it is easy
to calculate the effective Hamiltonian of the interaction between the
parallel troughs, mediated by the channeled tails (\ref{tail}), and thus
approximate the full 2D model (\ref{2DGPtwowell}), (\ref{twosquare1}) by the
system of coupled 1D equations with the local coupling:
\begin{eqnarray}
i\partial _{t}\psi _{1} &=&-\frac{1}{2}\partial _{y}^{2}\psi _{1}-|\psi
_{1}|^{2}\psi _{1}-\kappa \delta (y)\psi _{2},  \notag \\
&&  \label{1D} \\
i\partial _{t}\psi _{2} &=&-\frac{1}{2}\partial _{y}^{2}\psi _{2}-|\psi
_{2}|^{2}\psi _{2}-\kappa \delta (y)\psi _{1},  \notag
\end{eqnarray}%
with the coupling coefficient identified by equating the above-mentioned 2D
interaction Hamiltonian to its counterpart corresponding to the 1D system (%
\ref{1D}):
\begin{equation}
\kappa =\sqrt{\pi ^{2}-2\alpha _{1}\alpha _{3}^{2}}\exp \left( -\frac{\alpha
_{2}}{\alpha _{3}}\sqrt{\pi ^{2}-2\alpha _{1}\alpha _{3}^{2}}\right) .
\label{kappa}
\end{equation}%
In fact, by an additional rescaling of Eqs. (\ref{1D}) it is possible to fix
$\kappa \equiv 1$, which is adopted below, in Section IV. To understand the
genericity of the results produced by the 1D system with the $\delta $%
-functional coupling profile, we will also consider the 1D system with the $%
\delta $-function replaced by the Gaussian profile with a finite width, $%
y_{0},$%
\begin{equation}
\delta (y)\rightarrow \exp \left( -y^{2}/y_{0}^{2}\right) ,  \label{Gauss}
\end{equation}%
keeping the coefficient in front of the Gaussian to be $1$.

\section{Symmetric and asymmetric two-dimensional solitons}

Stationary soliton solutions to Eqs. (\ref{2DGPtwowell}), (\ref{twosquare1})
were found by means of the imaginary-time integration method \cite{imaginary}%
. The stability of the so generated solitons was then tested by direct
simulations of the perturbed evolution in real time. Typical examples of
stable symmetric and asymmetric 2D solitons are shown in Fig. \ref%
{kindsoliton}.

\begin{figure}[tbp]
\begin{center}
\includegraphics[width=8 cm]{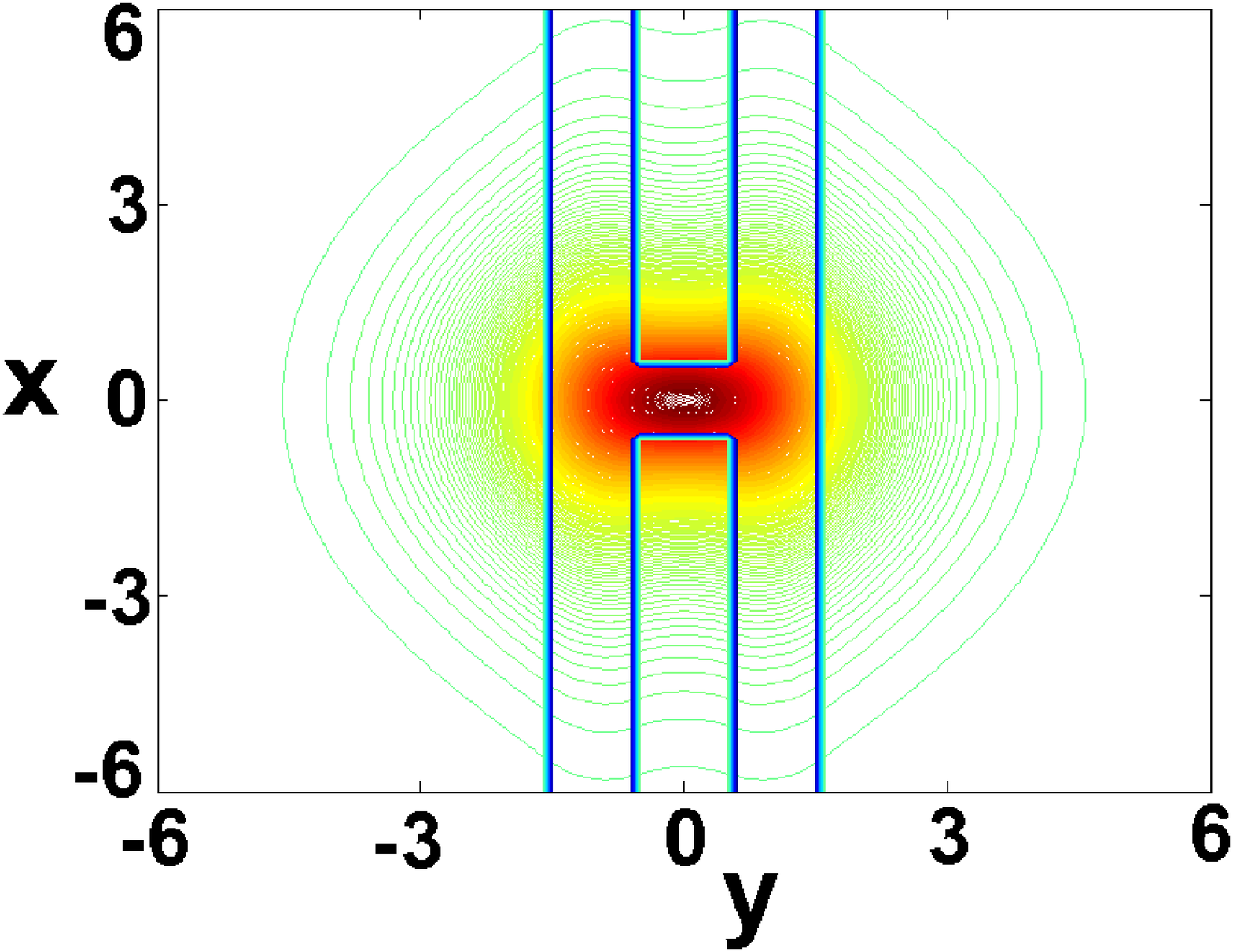} \includegraphics[width=8 cm]{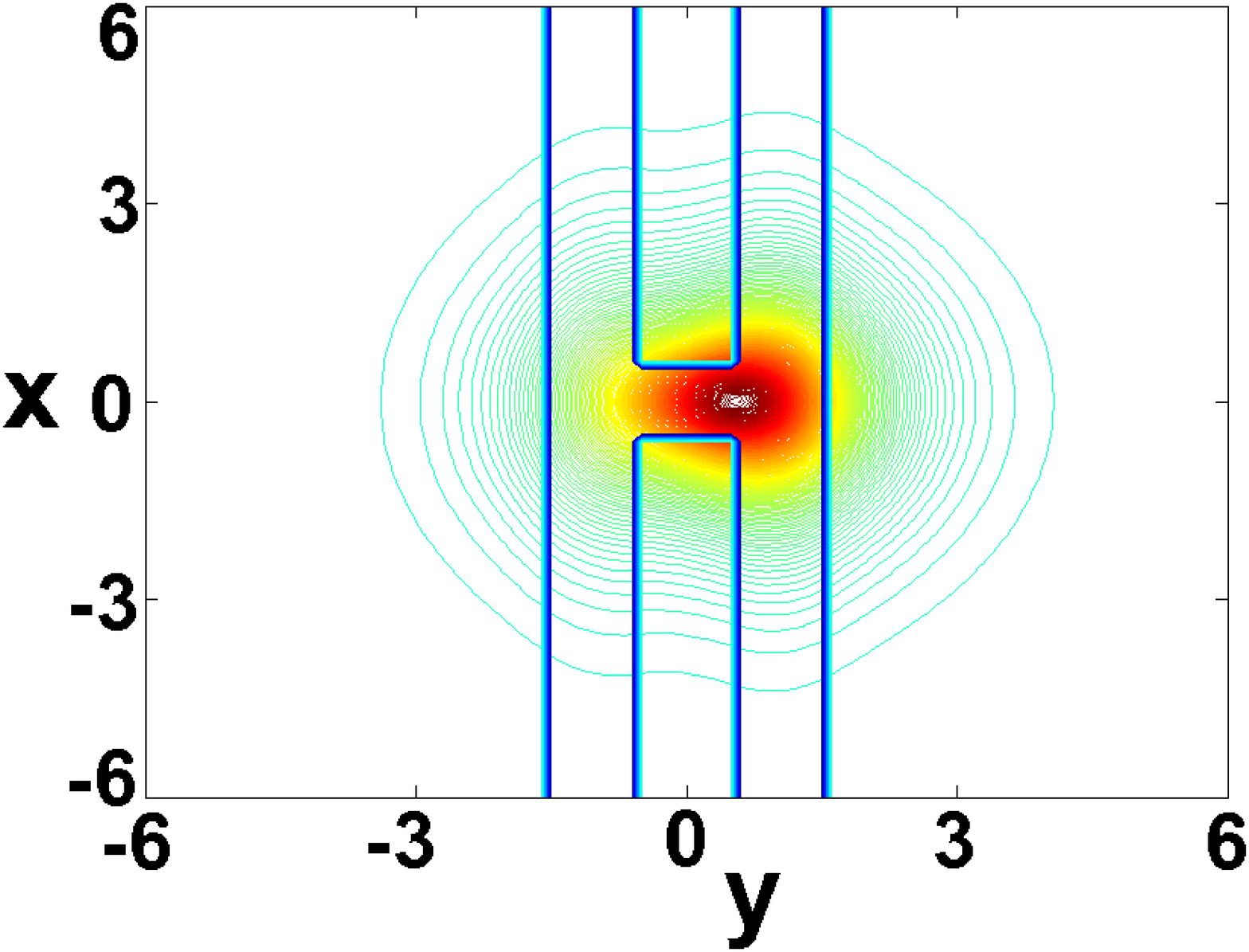}
\end{center}
\caption{(Color online) Examples of stable 2D symmetric (left) and
asymmetric (right) solitons. The respective values of the scaled norm are $%
N_{\mathrm{2D}}=4.05$ and $N_{\mathrm{2D}}=4.5$, respectively. The other
parameters are $\protect\alpha _{1}=\protect\alpha _{2}=\protect\alpha %
_{3}=1 $.}
\label{kindsoliton}
\end{figure}

Varying the set of control dimensionless parameters, $\left( \alpha
_{1},\alpha _{2},\alpha _{3}\right) $, we identified the critical value, $N_{%
\mathrm{cr}}$, of the scaled norm $N_{\mathrm{2D}}$, at which the SBB occurs
and asymmetric states appear. The natural measure for the asymmetry of such
states is defined as
\begin{equation}
\nu \equiv \frac{\int_{-\infty }^{+\infty }{d}y\left[ \int_{0}^{\infty }{d}%
x|\Psi (x,y)|^{2}-\int_{-\infty }^{0}{d}x|\Psi (x,y)|^{2}\right] }{N_{%
\mathrm{2D}}}.  \label{asym11}
\end{equation}

\begin{figure}[tbp]
\begin{center}
\includegraphics[width=8.5cm]{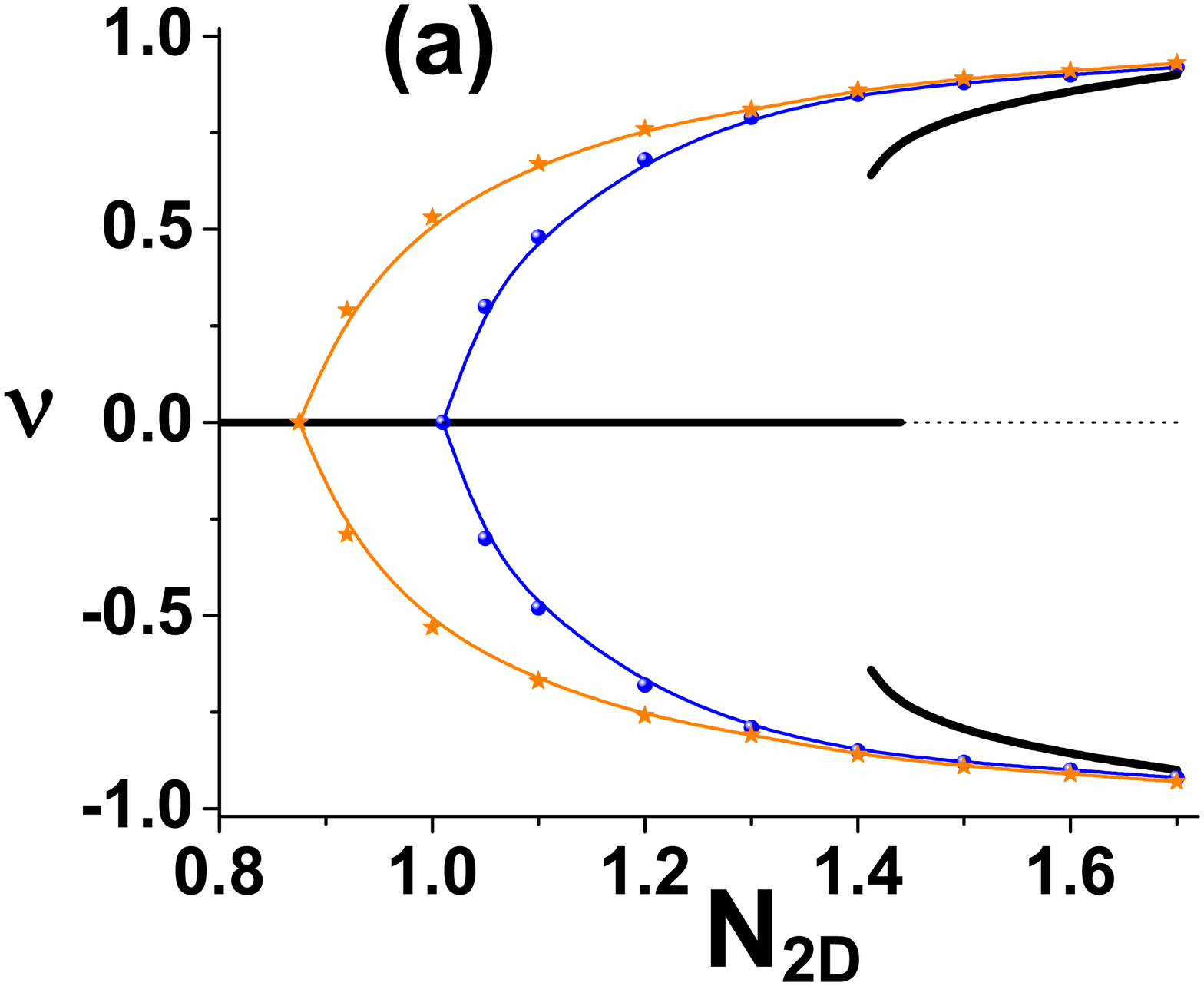} %
\includegraphics[width=8.5cm]{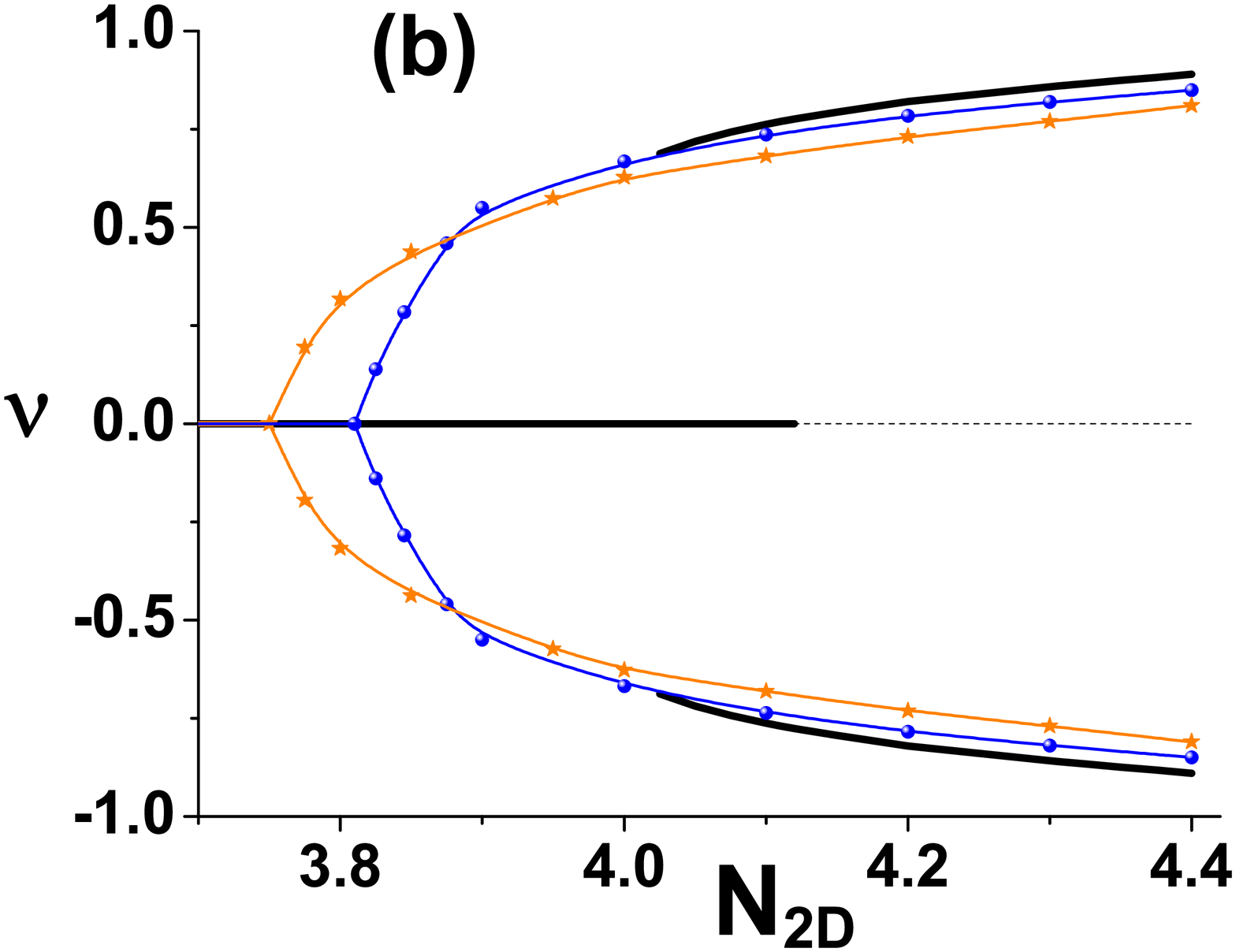}
\end{center}
\caption{(Color online) Bifurcation diagrams showing the asymmetry of the 2D
solitons as a function of the total norm. The parameters are $\protect\alpha %
_{1}=1,\protect\alpha _{2}=4$ in (a), and $\protect\alpha _{1}=\protect%
\alpha _{2}=1$ in (b) (fixed depth of the parallel troughs and the distance
between them), while the scaled width of the transverse rung is varying: $%
\protect\alpha _{3}=0.5$ (orange curves in each panel), $\protect\alpha %
_{3}=0.25$ (blue curves) and $\protect\alpha _{3}=0$ (black curves). In the
case of $\protect\alpha _{3}=0$ (no transverse linkage through the rung),
the bifurcation is subcritical, in agreement with Ref. \protect\cite{Warsaw1}
(only stable portions of the subcritical diagram are displayed in this
case). In other cases, the character of the bifurcation is clearly
supercritical.}
\label{diagram2d11}
\end{figure}

As mentioned in the Introduction, the 2D system without the transverse rung,
which is tantamount to the present 2D model with $W=\alpha _{3}=0$ \cite%
{Warsaw1}, as well as its 1D counterpart \cite{Arik}, based on the system of
1D equations uniformly coupled by linear terms, feature the SBB of the
subcritical type. The difference of the present model is that, with the
increase of $\alpha _{3}$ from zero, the character of the bifurcation
quickly switches to supercritical (in other words, the kind of the
respective phase transition switches from first to second), as shown in
detail by means of the set of bifurcation diagrams in Fig. \ref{diagram2d11}%
, where control parameters $\alpha _{1}$ are $\alpha _{2}$ are fixed, while $%
\alpha _{3}$ is varied. In terms of the underlying system, this means the
gradual increase of the width of the transverse rung in Fig. \ref%
{potenhshaped}. As verified by direct simulations (not shown here in
detail), all the solution branches displayed in Fig. \ref{diagram2d11} are
dynamically stable.

The transition to the supercritical bifurcation is a consequence of the
enhancement of the local transverse coupling between the troughs through the
rung. A qualitatively similar change of the bifurcation was recently
observed in dual discrete chains, with the transition from the uniform
transverse linear coupling \cite{Herring} to that at a single site \cite%
{Ljupco}.

\begin{figure}[tbp]
\begin{center}
\includegraphics[width=8cm]{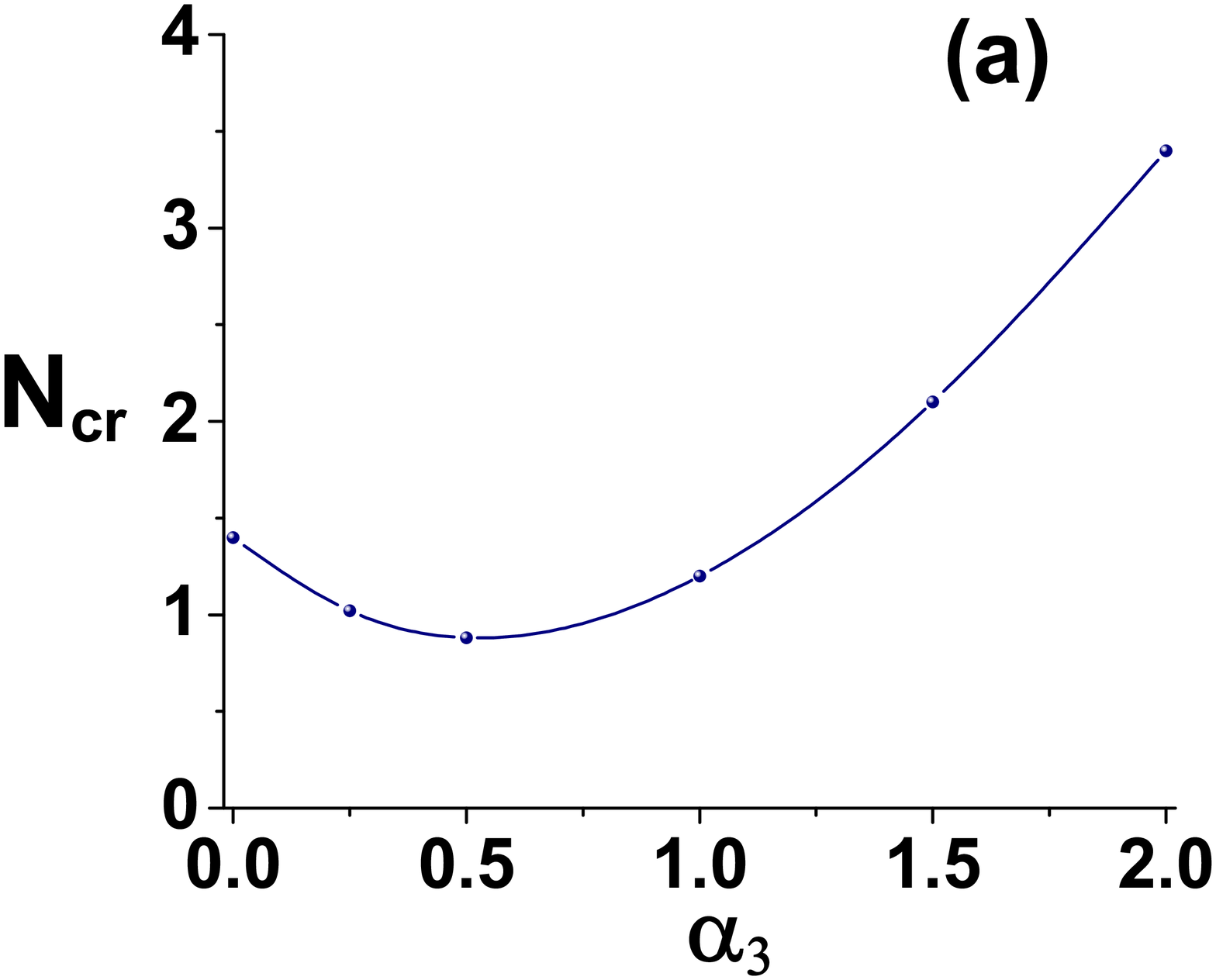}
\includegraphics[width=8cm]{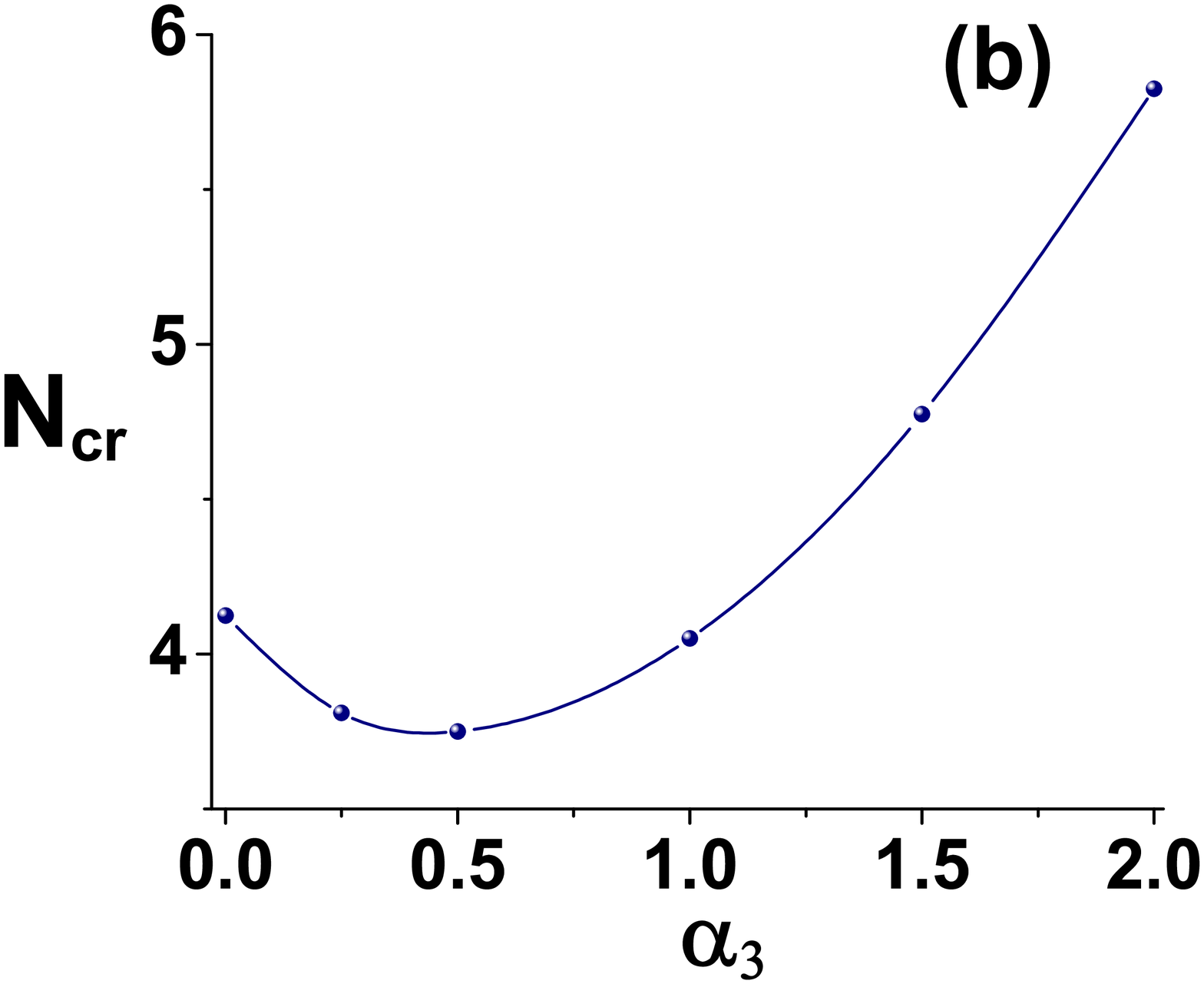} %
\end{center}
\caption{(Color online) The total norm of the wave field at the point of the
symmetry-breaking bifurcation, in the 2D model, versus the relative width of
the transverse rung, $\protect\alpha _{3}$. The other parameters are $%
\protect\alpha _{1}=1,\protect\alpha _{2}=4$ (a) and $\protect\alpha _{1}=%
\protect\alpha _{2}=1$ (b).}
\label{alpha3a}
\end{figure}

\begin{figure}[tbp]
\begin{center}
\includegraphics[width=12cm]{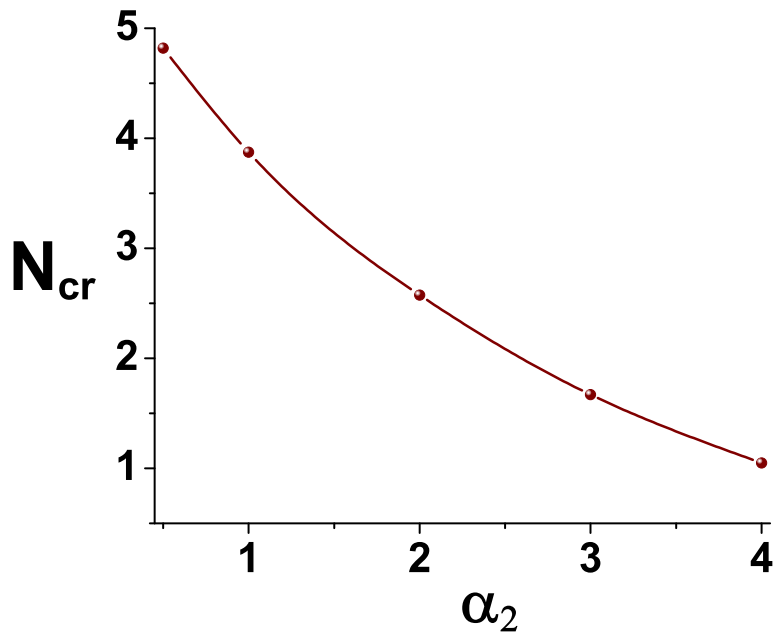}
\end{center}
\caption{(Color online) The norm at the point of the symmetry-breaking
bifurcation in the 2D model versus the relative separation between the
parallel troughs, $\protect\alpha _{2}$. The other parameters are $\protect%
\alpha _{1}=1,\protect\alpha _{3}=0.25$.}
\label{alpha2}
\end{figure}

The bifurcation picture is further characterized, in Figs. \ref{alpha3a} and %
\ref{alpha2}, respectively, by dependences of the value of the norm at the
bifurcation point, $N_{\mathrm{cr}}$, on the relative width of the
transverse rung, $\alpha _{3}$, and the relative distance between the
troughs, $\alpha _{2}$. Because, as said above, the increase of $\alpha _{3}$
(making the rung wider) implies the strengthening of the linear coupling
between the parallel troughs, one should expect the growth of $N_{\mathrm{cr}%
}$ with $\alpha _{3}$. This is, generally, observed in Fig. \ref{alpha3a},
but after an initial \emph{decrease}. This surprising feature is explained
by the above-mentioned change of the character of the bifurcation, from
subcritical to supercritical. On the other hand, the monotonous decrease of $%
N_{\mathrm{cr}}$ with the increase of $\alpha _{2}$ is a natural consequence
of the weakening strength of the coupling between the toughs.

The coupling between the troughs is, obviously, sensitive to the relative
width of the separating barrier, with respect to the troughs, which is
measured by $\alpha _{2}$, see Fig. \ref{potenhshaped} and Eq. (\ref{alpha}%
). The fact that the bifurcation diagrams, and the respective critical
values of the norm, are quite similar in panels (a) and (b) in Figs. \ref%
{diagram2d11} and \ref{alpha3a}, which pertain to $\alpha _{2}=4$ and $%
\alpha _{2}=1$, respectively, demonstrates that the type of the symmetry
breaking reported in this section comprises a broad parametric area.

Finally, the simulations demonstrate that the 2D solitons suffer the
collapse, in the present model, exactly when it is expected, i.e., when the
total scaled norm exceeds the well-known threshold value, $N_{\mathrm{thr}%
}\approx 5.85$ \cite{Berge}.

\section{The one-dimensional system: exact, variational, and numerical
solutions}

\subsection{Exact solutions for the linear coupling with the
delta-functional profile}

\subsubsection{General analysis}

A remarkable feature of the 1D system based on Eq. (\ref{1D}) with the
transverse coupling accounted for by the $\delta $-functions is a
possibility to find exact stationary solutions for the solitons of all the
types, symmetric, antisymmetric, and asymmetric (note that exact solutions
are not available in the discrete counterpart of the system considered in
Ref. \cite{Ljupco}). Stationary solutions to Eq. (\ref{1D}) with a given
(negative) chemical potential, $\mu $ (in the application to optics, $-\mu $
is the propagation constant), are looked for as
\begin{equation}
\left\{ \psi _{1}\left( y,t\right) ,\psi _{2}\left( y,t\right) \right\}
=e^{-i\mu t}\left\{ u(y),v(y)\right\} ,  \label{12}
\end{equation}%
where real functions $u(y)$ and $v(y)$ obey the following equations:%
\begin{eqnarray}
\mu u &=&-\frac{1}{2}\frac{d^{2}u}{dy^{2}}-u^{3}-\delta (y)v,  \notag \\
&&  \label{uv} \\
\mu v &=&-\frac{1}{2}\frac{d^{2}v}{dy^{2}}-v^{3}-\delta (y)u  \notag
\end{eqnarray}%
[recall $\kappa \equiv 1$ is fixed in Eq. (\ref{1D}) by means of rescaling].
As follows from Eqs. (\ref{uv}), at $y=0$ the solutions must satisfy the
following boundary conditions:%
\begin{equation}
\Delta \left( u^{\prime }\right) =-2v,~\Delta \left( v^{\prime }\right) =-2u,
\label{Delta}
\end{equation}%
where $\Delta $ stands for the jump of the derivative at $y=0$. Solutions to
Eqs. (\ref{uv}) and (\ref{Delta}) are looked for as%
\begin{equation}
u=\eta ~\mathrm{sech}\left( \eta \left( |y|~+a\right) \right) ,~v=s\eta ~%
\mathrm{sech}\left( \eta \left( |y|~+b\right) \right) ,~\eta \equiv \sqrt{%
-2\mu },  \label{sech}
\end{equation}%
where $s=+1$ and $s=-1$ correspond to the symmetric and antisymmetric
states, respectively (or their asymmetric deformations), while shifts $a>0$
and $b>0$ are determined by equations following from the substitution of
ansatz (\ref{sech}) into Eqs. (\ref{Delta}):%
\begin{eqnarray}
\eta \frac{\sinh \left( \eta a\right) }{\cosh ^{2}\left( \eta a\right) } &=&%
\frac{s}{\cosh \left( \eta b\right) },  \notag \\
&&  \label{ab} \\
\eta \frac{\sinh \left( \eta b\right) }{\cosh ^{2}\left( \eta b\right) } &=&%
\frac{s}{\cosh \left( \eta a\right) }.  \notag
\end{eqnarray}%
Remind $\eta \equiv \sqrt{-2\mu }$ is treated here as an arbitrary constant
parameterizing the family of solutions.

\subsubsection{Symmetric and antisymmetric solutions}

Symmetric modes correspond to $s=+1$ and $a=b$, in which case Eq. (\ref{ab})
yields%
\begin{equation}
\exp \left( 2\eta a_{\mathrm{symm}}\right) =\left( \eta +1\right) /\left(
\eta -1\right) .  \label{symm}
\end{equation}%
As seen from here, the symmetric solution exists for $\eta >1$, i.e., $\mu
<-1/2$, with the total norm which can be readily calculated:
\begin{equation}
N=\int_{-\infty }^{+\infty }\left[ u^{2}(y)+v^{2}(y)\right] dy\equiv
N_{u}+N_{v}=4\left( \eta -1\right) .  \label{N}
\end{equation}

Antisymmetric states, with $s=-1$ and $a=b<0$, are also possible, with $|a|$
given by the same expression (\ref{symm}) as in the symmetric case. However,
the antisymmetric states are expected to be completely unstable, as they
correspond to a maximum, rather than minimum, of the respective interaction
Hamiltonian, therefore they are not be considered below (the instability of
the antisymmetric states is corroborated by numerical tests, see, e.g., Fig. %
\ref{AntiSymmetric2} below).

\subsubsection{Asymmetric solitons}

For the most interesting asymmetric solutions with $a\neq b$ and $s=+1$, Eq.
(\ref{ab}) can be transformed into the following system:%
\begin{eqnarray}
\tanh ^{2}\left( \eta a\right) +\tanh ^{2}\left( \eta b\right) &=&1-\eta
^{-2}, \\
\tanh \left( \eta a\right) \cdot \tanh \left( \eta b\right) &=&\eta ^{-2}~,
\end{eqnarray}%
an explicit solution to which is%
\begin{eqnarray}
\tanh \left( \eta a\right) &=&\frac{1}{2}\left( \sqrt{1+\eta ^{-2}}-\sqrt{%
1-3\eta ^{-2}}\right) ,  \notag \\
&&  \label{tanhtanh} \\
\tanh \left( \eta b\right) &=&\frac{1}{2}\left( \sqrt{1+\eta ^{-2}}+\sqrt{%
1-3\eta ^{-2}}\right) .  \notag
\end{eqnarray}%
As seen from these expressions, with the increase of $\eta $, i.e., with the
growth of the norm of the symmetric solutions [see Eq. (\ref{N})], the
asymmetric modes emerge and exist at%
\begin{equation}
\eta \geq \eta _{0}\equiv \sqrt{3},  \label{0}
\end{equation}%
with the following values of the norm in the two troughs:%
\begin{eqnarray}
N_{u} &=&2\eta -\sqrt{\eta ^{2}+1}+\sqrt{\eta ^{2}-3},  \notag \\
&&  \label{Nuv} \\
N_{v} &=&2\eta -\sqrt{\eta ^{2}+1}-\sqrt{\eta ^{2}-3}.  \notag
\end{eqnarray}%
Accordingly, the total norm of the asymmetric solution is%
\begin{equation}
N\equiv N_{u}+N_{v}=2\left( 2\eta -\sqrt{\eta ^{2}+1}\right) ,  \label{total}
\end{equation}%
or, inversely, constant $\eta $ may be expressed in terms of the total norm,
which is then treated as the intrinsic parameter of the family of the
asymmetric solitons:%
\begin{equation}
\eta =\frac{1}{3}\left( N+\sqrt{\frac{1}{4}N^{2}+3}\right) .  \label{eta}
\end{equation}%
The corresponding asymmetry measure is%
\begin{equation}
\nu \equiv \frac{N_{u}-N_{v}}{N_{u}+N_{v}}=\frac{\sqrt{\eta ^{2}-3}}{2\eta -%
\sqrt{\eta ^{2}+1}}~.
\end{equation}%
Substituting here $\eta $ from Eq. (\ref{eta}), we obtain an eventual
expression for the asymmetry parameter as a function of the total norm:
\begin{equation}
\nu =\frac{\sqrt{\left( N^{2}-4\right) ^{2}-12\left( 2N-\sqrt{N^{2}+12}%
\right) ^{2}}}{2\left( N^{2}-4\right) -\sqrt{\left( N^{2}-4\right)
^{2}+4\left( 2N-\sqrt{N^{2}+12}\right) ^{2}}}~,  \label{final}
\end{equation}%
which provides a full analytical description of the SBB in the 1D system.
The norm of the asymmetric solitons assumes values $N\geq N_{0}=4\left(
\sqrt{3}-1\right) \approx 2.93$, with $N_{0}$ corresponding to the
symmetry-breaking point (\ref{0}).

\begin{figure}[tbp]
\begin{center}
\includegraphics[width=10cm]{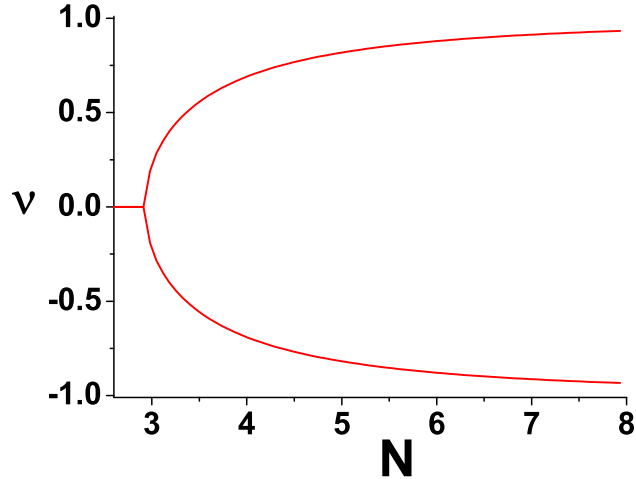}
\end{center}
\caption{(Color online) The bifurcation diagram produced by the exact
solution of the 1D system with the $\protect\delta $-functional profile of
the linear coupling.}
\label{diagramdirac1}
\end{figure}

The bifurcation diagram predicted by Eq. (\ref{final}), i.e., $\nu $ as a
function of $N$, is displayed in Fig. \ref{diagramdirac1}. Obviously, the
bifurcation revealed by the exact solution is supercritical, being quite
similar to the supercritical SBB in the 2D model, which is displayed above
in Fig. \ref{diagram2d11}. Direct simulations (with the ideal $\delta $%
-functions replaced by their regularized versions, see below) confirm that,
as expected, all the solution branches shown in Fig. \ref{diagramdirac1} are
stable.

\subsection{The coupling with the Gaussian profile}

To check how generic the exact solution found with the $\delta $-functional
coupling profile is, it is natural to compare the results to those generated
by the Gaussian profile (\ref{Gauss}). Because exact solutions are not
available in this case, we will tackle the problem by means of the VA
(variational approximation), which is an effective tool for the analysis of
a broad class of systems similar to the present one \cite{VA}.

To apply the VA, we use the expression for the energy of the 1D system (\ref%
{1D}), (\ref{Gauss}):
\begin{eqnarray}
E &=&\int_{-\infty }^{+\infty }dy\left[ \frac{1}{2}\sum_{j=1}^{2}\left\vert
\partial _{y}\psi _{j}\right\vert ^{2}-\frac{1}{2}\sum_{j=1}^{2}|\psi
_{j}|^{4}\right.  \notag \\
&&\left. -\exp \left( -\frac{y^{2}}{y_{0}^{2}}\right) \left( \psi _{1}\psi
_{2}^{\ast }+\psi _{1}^{\ast }\psi _{2}\right) \right] ,  \label{E}
\end{eqnarray}%
and introduce the following ansatz (hereafter we use $\pm $ instead of
subscripts 1 and 2):
\begin{equation}
\psi _{\pm }(y,t)=e^{-i\mu t}\sqrt{\frac{N\left( 1\pm \nu \right) }{2W\sqrt{%
\pi }}}\exp \left( -\frac{y^{2}}{2W^{2}}\right) ,  \label{exp}
\end{equation}%
where $W$ is the width of the soliton, and $\nu $ is the asymmetry parameter
defined as
\begin{equation}
\nu =\frac{1}{N}\int_{-\infty }^{+\infty }dy\,\left( |\psi _{1}|^{2}-|\psi
_{2}|^{2}\right) ,  \label{z}
\end{equation}%
cf. Eq. (\ref{asym11}). Substituting ansatz (\ref{exp}) into expression (\ref%
{E}) yields the energy as a function of variational parameters, $W$ and $\nu
$:
\begin{equation}
E=\frac{N}{4W^{2}}-\frac{Ny_{0}\sqrt{1-\nu ^{2}}}{\sqrt{y_{0}^{2}+W^{2}}}-%
\frac{N^{2}(1+\nu ^{2})}{4W\sqrt{2\pi }}.  \label{funcener1}
\end{equation}

The variational equations following from expression (\ref{funcener1}), $%
\partial E/\partial W=\partial E/\partial \nu =0$, take the following form:
\begin{eqnarray}
\frac{1}{W^{4}}-\frac{2y_{0}\sqrt{1-\nu ^{2}}}{(y_{0}^{2}+W^{2})^{3/2}}-%
\frac{N(1+\nu ^{2})}{2\sqrt{2\pi }W^{3}} &=&0,  \notag \\
&&  \label{Wnu} \\
\nu \left( \frac{y_{0}}{\sqrt{(y_{0}^{2}+W^{2})(1-\nu ^{2})}}-\frac{N}{2%
\sqrt{2\pi }W}\right) &=&0.  \notag
\end{eqnarray}%
These equations were solved in a numerical form, to find parameters $W$ and $%
\nu $ of stationary solutions as functions of the free parameters, $y_{0}$
and $N$. The stability of the solutions was verified, in the framework of
the VA, by checking if they correspond to a local minimum of energy (\ref%
{funcener1}).

\begin{figure}[tbp]
\begin{center}
\includegraphics[width=8cm]{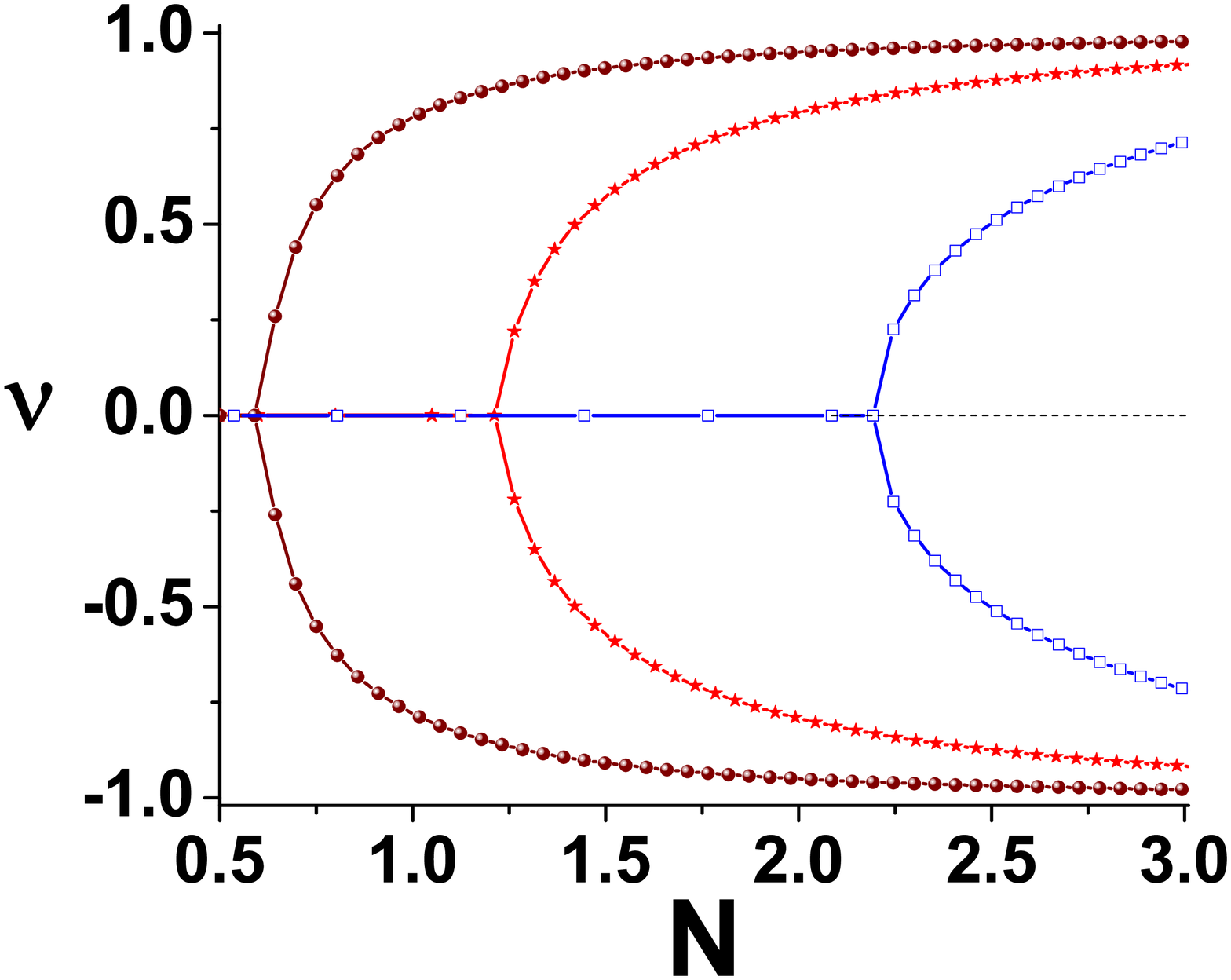} %
\includegraphics[width=8cm]{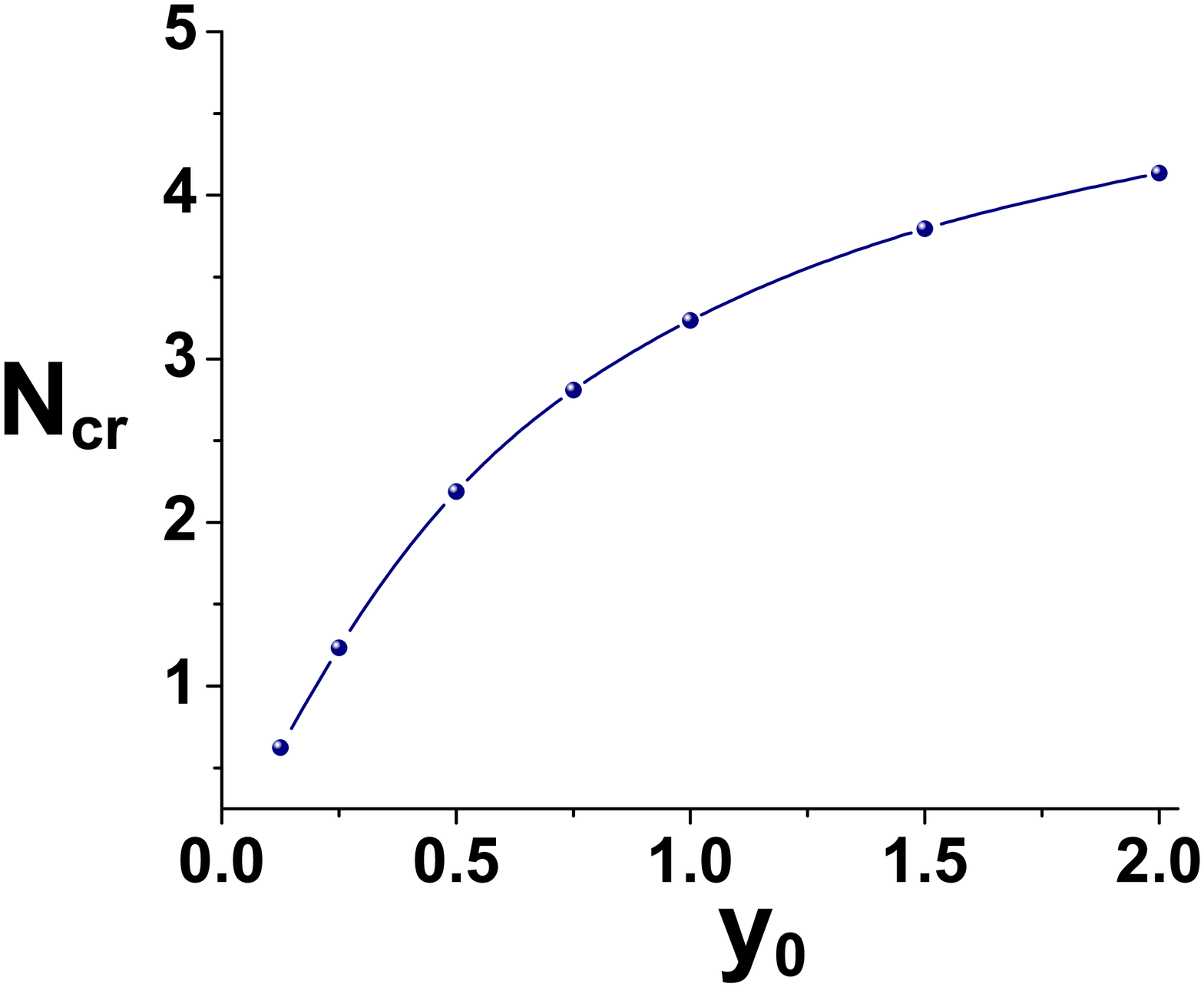}
\end{center}
\caption{(Color online) The bifurcation diagram produced by the 1D system
with the Gaussian profile of the coupling, Eqs. (\protect\ref{1D}) and (%
\protect\ref{Gauss}), in the framework of the variational
approximation. The width of the Gaussian profile is $y_{0}=0.5$
(squares), $0.25$ (stars) and $0.125$ (balls), respectively.}
\label{diagramgausian1}
\end{figure}

The so generated bifurcation diagrams are displayed in the left panel of
Fig. \ref{diagramgausian1} for several fixed values of width $y_{0}$ of the
Gaussian profile of the coupling. It is clearly seen that the SBB is
supercritical in this case too, being quite similar to that produced by the
exact solution for $\delta $-functional profile, cf. Fig. \ref{diagramdirac1}%
, and to the supercritical diagrams found in the 2D model, cf. Fig. \ref%
{diagram2d11}. The respective dependence of the norm at the bifurcation
point on the width of the Gaussian profile, $N_{\mathrm{cr}}(y_{0})$. In
fact, the latter plot is a 1D counterpart of the one which, in the 2D model,
shows the dependence of $N_{\mathrm{cr}}$ on the relative width of the
transverse rung, $\alpha _{3}$, cf. Fig. \ref{alpha3a}.

We have also constructed stationary states of the 1D system with the
Gaussian-shaped coupling as numerical solutions of Eqs. (\ref{1D}) and (\ref%
{Gauss}). Two different algorithms were used for this purpose, the
imaginary-time integration and the Newton-Kantorovich iteration method, both
yielding identical results. Thus, three types of stationary configurations
were found---symmetric, antisymmetric, and asymmetric ones. Their stability
was tested by direct simulations of the evolution in real time. It has been
confirmed that the respective SBB is indeed supercritical, and the
variational results are very close to the numerically exact ones. It was
also found that antisymmetric modes (which do not undergo the bifurcation)
are unstable (as mentioned above), see an example in Fig. \ref%
{AntiSymmetric2}. The two-peak profile of the stationary mode in this figure
is a consequence of its antisymmetry (the Gaussian profile of the coupling
is effectively repulsive, in this case, which also explains the instability
of the mode). It is seen that almost the entire norm is spontaneously
transferred into a single trough, where the two peaks split into separating
single-component solitons.

\begin{figure}[tbp]
\begin{center}
\includegraphics[width=11cm]{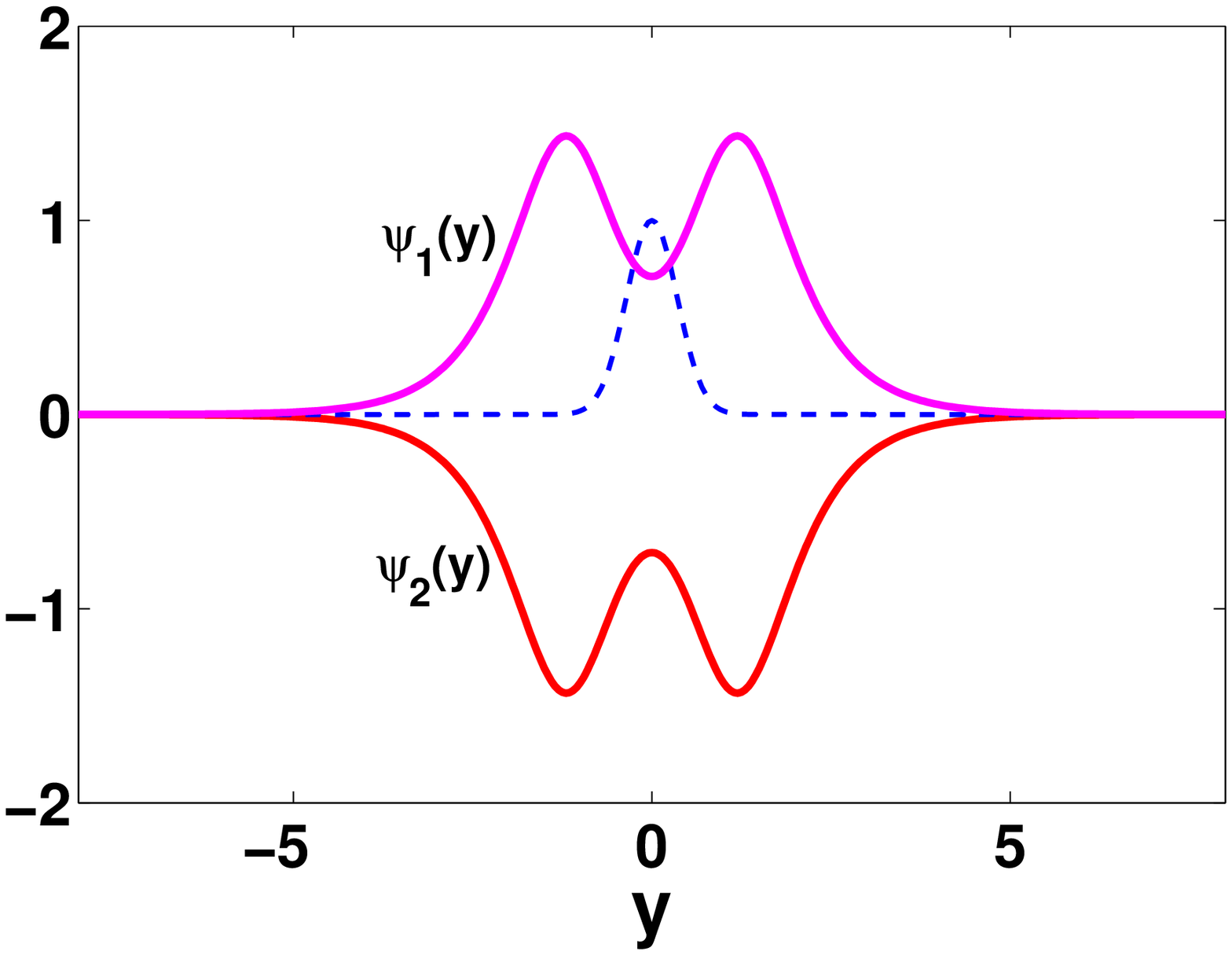} %
\includegraphics[width=12.6cm]{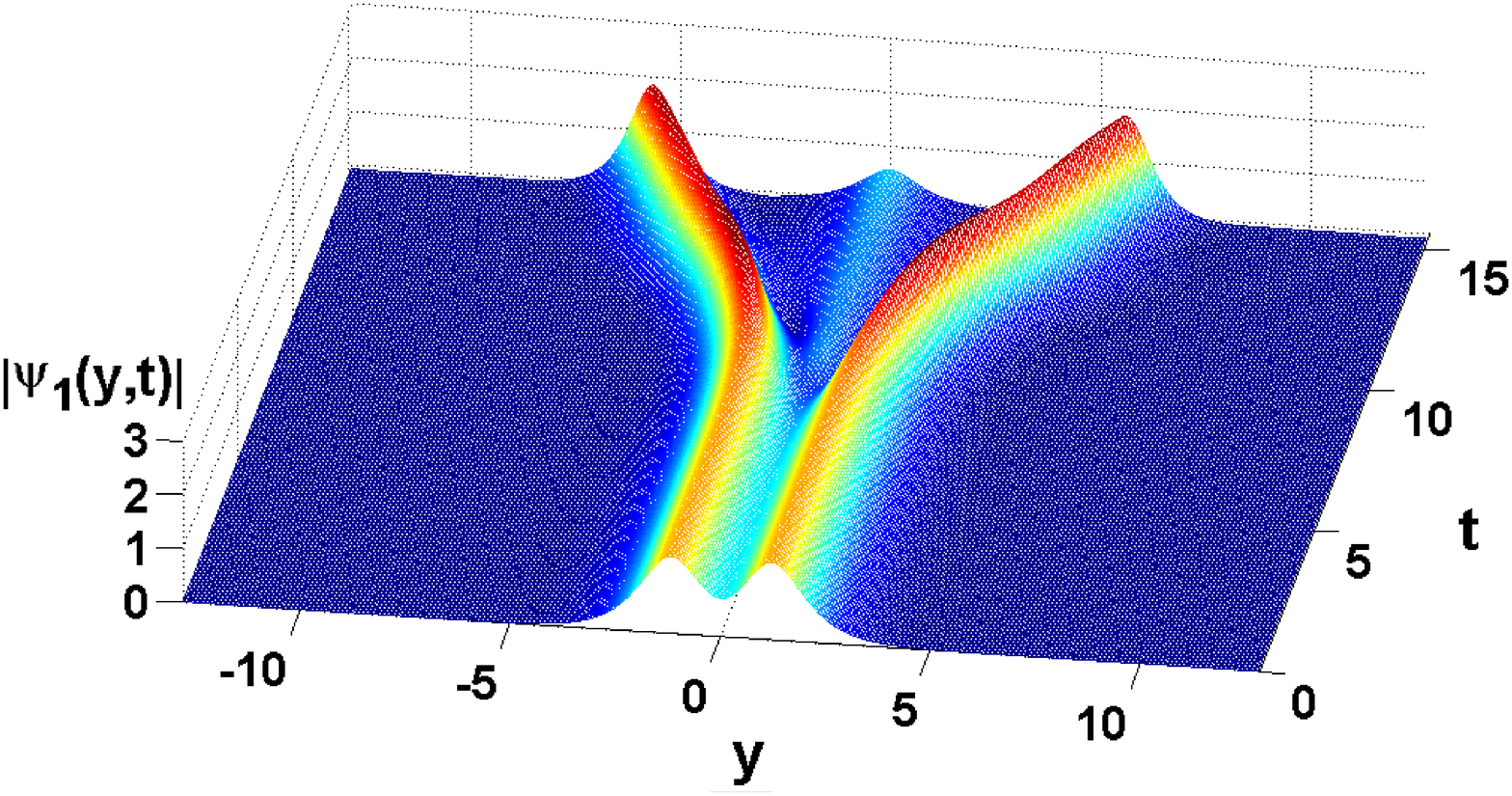} %
\includegraphics[width=12.6cm]{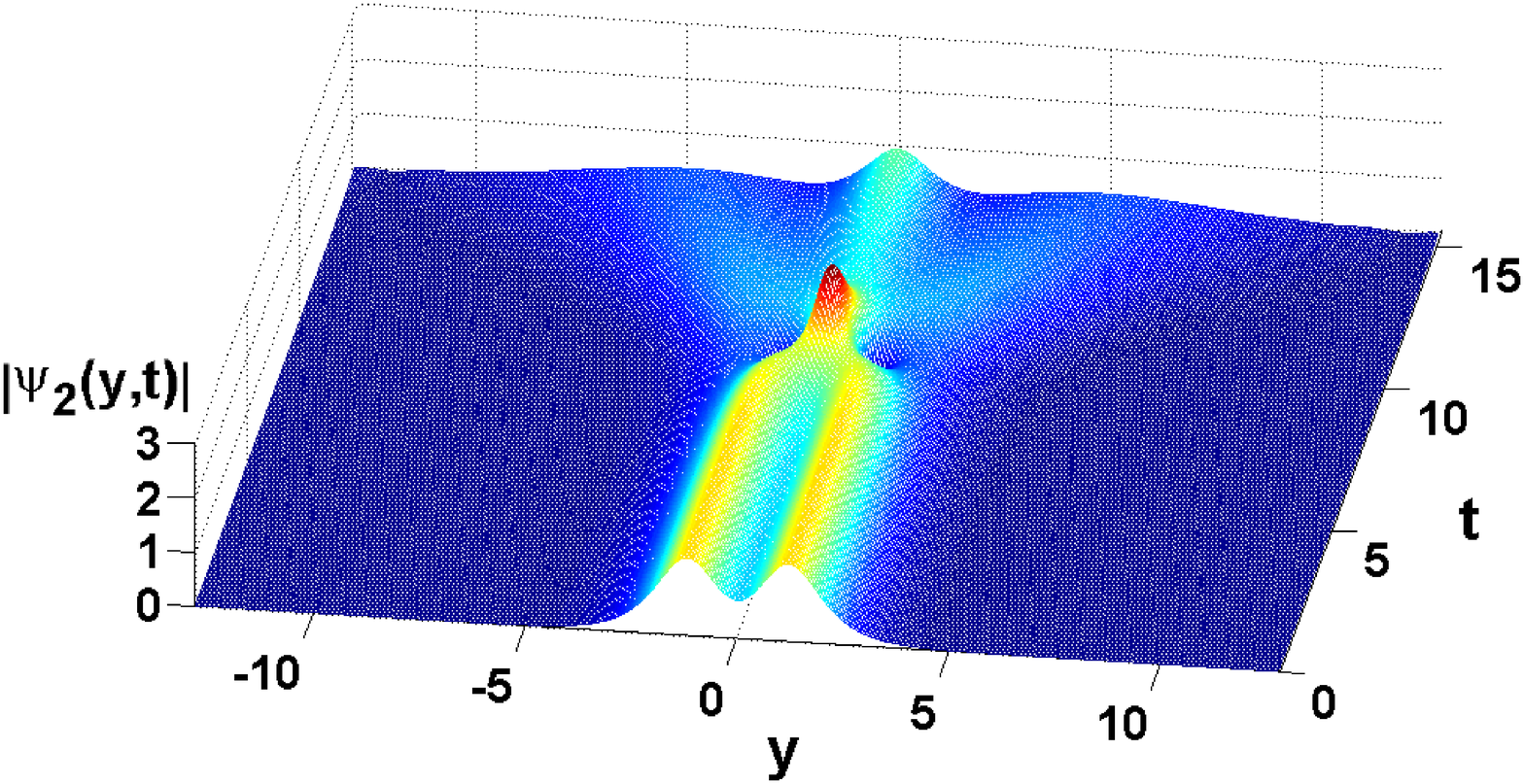}
\end{center}
\caption{(Color online) The shape of a typical antisymmetric solution and
its evolution in the 1D system (\protect\ref{1D}) with the Gaussian profile (%
\protect\ref{Gauss}) of the linear coupling. In the top panel, the
pink, red and (dash) blue curves display $\psi _{1}(y)$, $\psi
_{2}(y)$ in the stationary solution, and Gaussian profile $\exp
(-y^{2}/y_{0}^{2})$, respectively. The two lower panels show the
evolution of $\left\vert \protect\psi _{1}\left(
y,t\right) \right\vert $ and $\left\vert \protect\psi _{2}(y,t)\right\vert $%
, respectively. The parameters are: total norm $N=11.4$, chemical potential $%
\protect\mu =-1.03$, and the width of the Gaussian profile $y_{0}=0.5$.}
\label{AntiSymmetric2}
\end{figure}

We have also developed a similar analysis, using both the VA and numerical
solutions, for the 1D model with a rectangular profile of the transverse
coupling. The results (not shown here) are very similar to those generated
by the Gaussian profile.

\section{The combined linear-nonlinear coupling with the Gaussian profile}

\begin{figure}[tbp]
\begin{center}
\includegraphics[width=8cm]{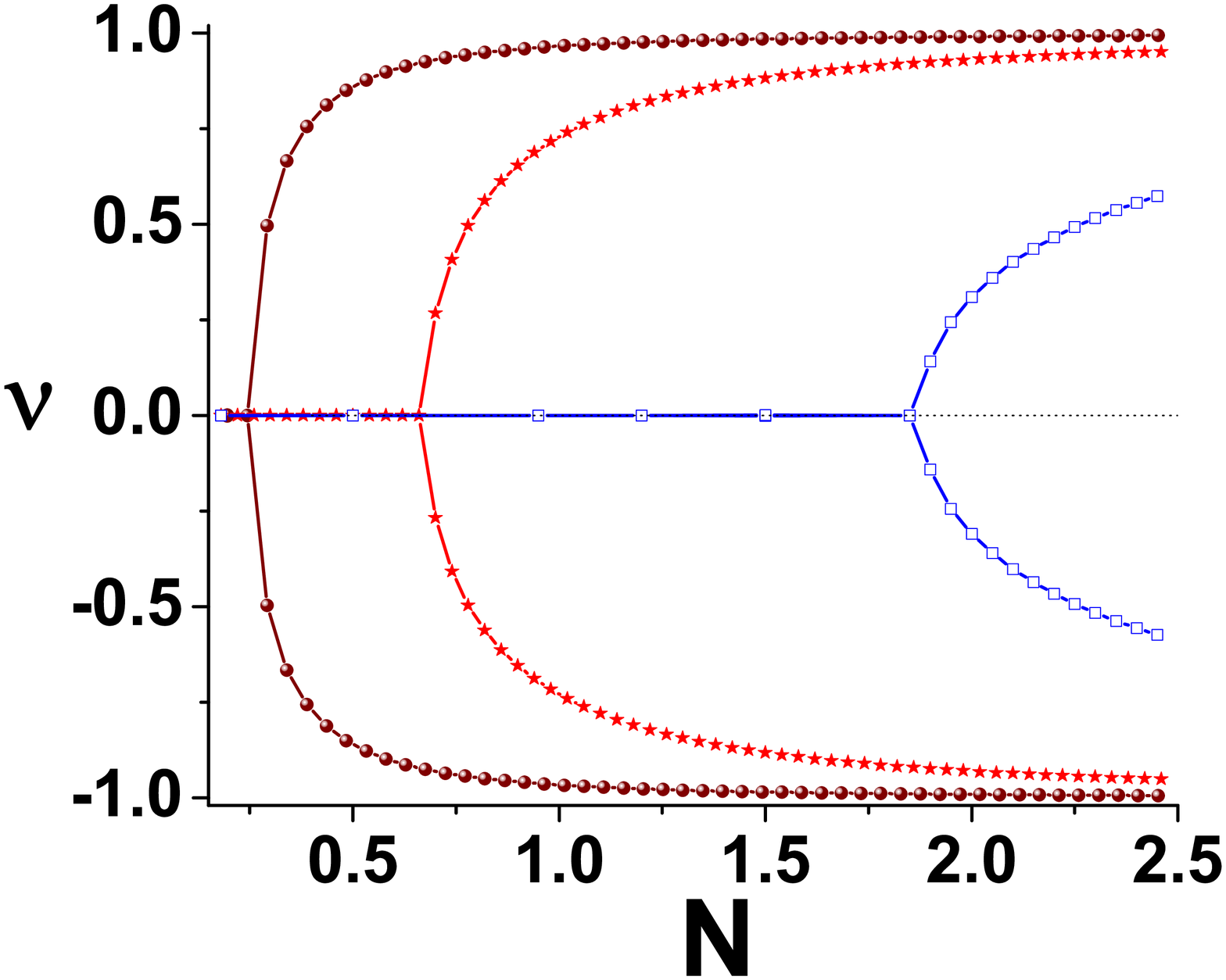} %
\includegraphics[width=8cm]{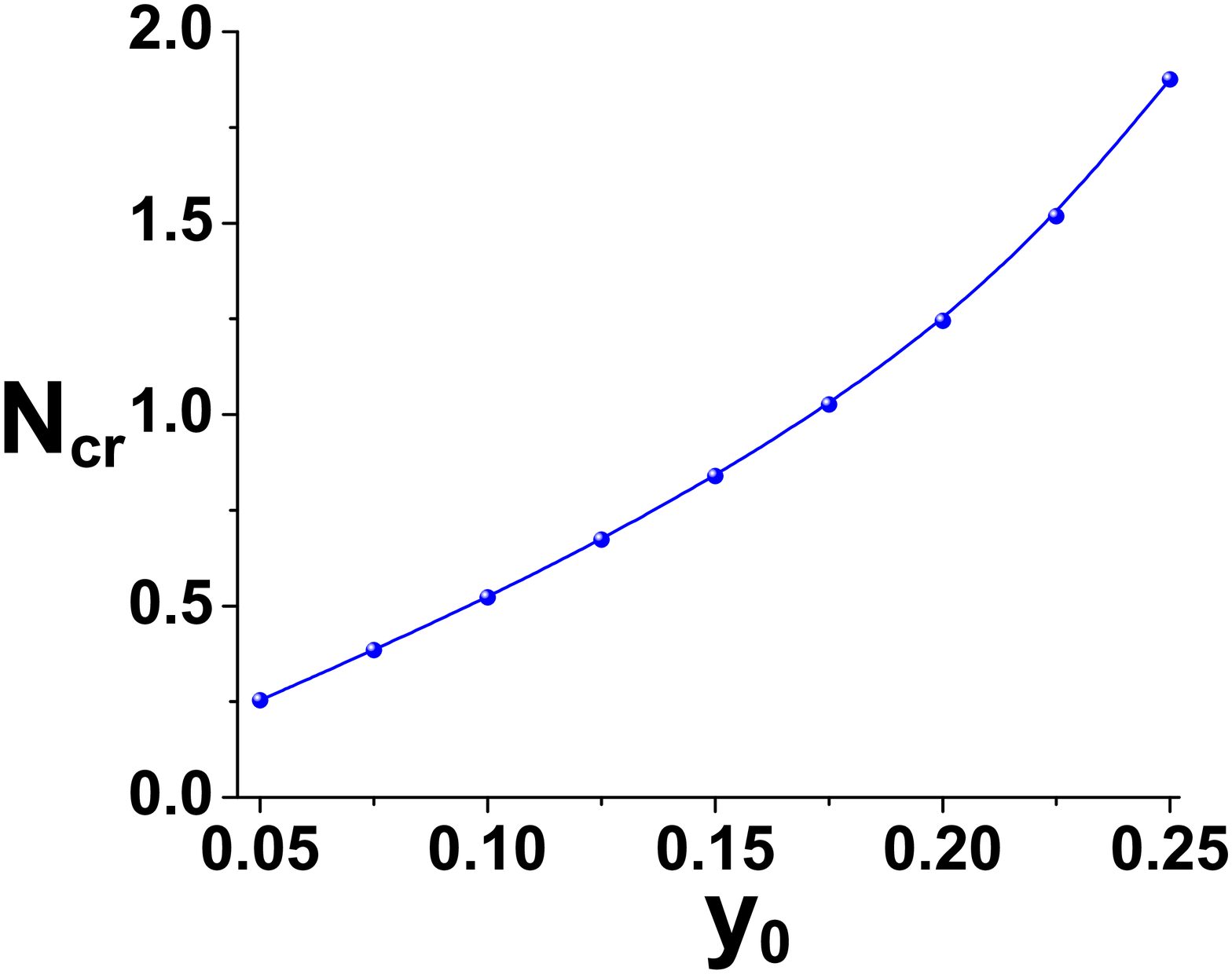}
\end{center}
\caption{(Color online) The same as in Figs. \protect\ref{diagramgausian1} ,
but for the 1D model with the Gaussian profile of the mixed linear-nonlinear
coupling, based on Eq. (\protect\ref{nonlin}) and the variational ansatz (%
\protect\ref{exp}). The width of the Gaussian profile is
$y_{0}=0.25$ (squares), $0.125$ (stars) and $0.05$ (balls),
respectively. The relative strength of the nonlinear and linear
couplings is $\protect\sigma _{0}=1$.} \label{diagramsnonlin1}
\end{figure}

It may also be relevant to take into regard a nonlinear correction to the
linear coupling between the two troughs, in the framework of the 1D system
(note that the full 2D model automatically takes into regard both linear and
nonlinear effects of the coupling). To this end, we adopt the following
modification of the 1D system (\ref{1D}), with the Gaussian coupling profile
(\ref{Gauss}):

\begin{eqnarray}
i\partial _{t}\psi _{1} &=&-\frac{1}{2}\partial _{y}^{2}\psi _{1}-|\psi
_{1}|^{2}\psi _{1}-\exp \left( -y^{2}/y_{0}^{2}\right) \left( \sigma |\psi
_{2}|^{2}\psi _{1}+\psi _{2}\right) ,  \notag \\
&&  \label{nonlin} \\
i\partial _{t}\psi _{2} &=&-\frac{1}{2}\partial _{y}^{2}\psi _{2}-|\psi
_{2}|^{2}\psi _{2}-\exp \left( -y^{2}/y_{0}^{2}\right) \left( \sigma |\psi
_{1}|^{2}\psi _{2}+\psi _{1}\right) ,  \notag
\end{eqnarray}%
where $\sigma $ is the relative strength of the nonlinear coupling. In this
case, the energy functional is
\begin{gather}
E=\int_{-\infty }^{+\infty }dy\left\{ \frac{1}{2}\sum_{j=1}^{2}\left\vert
\partial _{y}\psi _{j}\right\vert ^{2}-\frac{1}{2}\sum_{j=1}^{2}|\psi
_{j}|^{4}\right.  \notag \\
\left. -\exp \left( -\frac{y^{2}}{y_{0}^{2}}\right) \left[ \sigma |\psi
_{1}|^{2}|\psi _{2}|^{2}+\left( \psi _{1}\psi _{2}^{\ast }+\psi _{1}^{\ast
}\psi _{2}\right) \right] \right\} .  \label{Enonlin}
\end{gather}%
To apply the VA method, we adopt the same Gaussian ansatz (\ref{exp}) which
was used above. The substitution of the ansatz into functional (\ref{Enonlin}%
) yields
\begin{gather}
E=\frac{N}{4W^{2}}-Ny_{0}\sqrt{\frac{1-\nu ^{2}}{W^{2}+y_{0}^{2}}} \notag \\
-\frac{N^{2}}{4W\sqrt{\pi }}\left( \frac{\sigma y_{0}(1-\nu ^{2})}{\sqrt{%
W^{2}+2y_{0}^{2}}}+\frac{1+\nu ^{2}}{\sqrt{2}}\right) ,
\end{gather}%
the corresponding variational equations, $\partial E/\partial W=\partial
E/\partial \nu =0$, being
\begin{gather}
\frac{8y_{0}W\sqrt{1-\nu ^{2}}}{(W^{2}+y_{0}^{2})^{3/2}}+\frac{\sqrt{2}%
N(1+\nu ^{2})}{\sqrt{\pi }W^{2}}  \notag \\
+\frac{2\sigma y_{0}N(1-\nu ^{2})}{\sqrt{\pi (W^{2}+2y_{0}^{2})}}\left(
\frac{1}{W^{2}}+\frac{1}{W^{2}+2y_{0}^{2}}\right) -\frac{4}{W^{3}}=0,  \notag
\\ \label{LN1L}
\\
\nu \left[ \frac{2y_{0}}{\sqrt{(W^{2}+y_{0}^{2})(1-\nu ^{2})}}+\frac{N}{W%
\sqrt{\pi }}\left( \frac{\sigma y_{0}}{\sqrt{W^{2}+2y_{0}^{2}}}-\frac{1}{%
\sqrt{2}}\right) \right] =0.  \notag
\end{gather}%
The bifurcation diagram produced by the numerical solution of Eqs. (\ref{LN1L}%
), and the respective dependence of the total norm at the SBB point on $%
y_{0} $ are displayed (for $\sigma =1$) in Fig. \ref{diagramsnonlin1}.
\newline

The comparison of the dependences $N_{\mathrm{cr}}(y_{0})$, obtained
in the three versions of the1D model with the finite width of the
coupling profile (Gaussian, rectangular---which is not displayed
here in detail---and mixed linear-nonlinear) suggests that the
inclusion of the nonlinear coupling into Eq. (\ref{nonlin}) makes
this characteristic essentially more similar to its counterpart
obtained in the full 2D model, cf. Fig. \ref{alpha3a}: While in Fig.
\ref{diagramgausian1}, and in its counterpart obtained in the model
with the rectangular profile, the curves $N_{\mathrm{cr}}(y_{0})$
are convex, they are concave in both Figs. \ref{alpha3a} and \ref%
{diagramsnonlin1}. This conclusion is naturally explained by the
above-mentioned fact that the full 2D model automatically takes into account
both the linear and nonlinear coupling between the parallel troughs.

\section{Conclusion}

The objective of this work is to extend the study of the spontaneous
symmetry breaking in 1D and 2D solitons trapped in the systems of two
parallel tunnel-coupled potential troughs. Unlike recently studied models,
here we introduce the H-shaped potential, and focus on the SBB
(symmetry-breaking bifurcation) for solitons in this system. The analysis of
both the 2D and 1D versions of the model demonstrates that the transverse
link (the rung of the H-shaped potential profile) transforms the bifurcation
from subcritical into the supercritical one. In other words, the rung
controls the switching of the corresponding symmetry-breaking phase
transition from the first into second kind. A nontrivial manifestation of
the change of the SBB type is the non-monotonous dependence of the critical
value of the soliton's norm, at the bifurcation point, on the strength of
the transverse link: prior to the expected growth, it demonstrates a region
of the decrease. In the full 2D model, the results were obtained in the
numerical form. On the other hand, the 1D version with the $\delta $%
-functional profile of the local linear coupling between the troughs admits
the exact analytical solutions for the solitons of all the
types---symmetric, antisymmetric, and asymmetric---and, accordingly, the
bifurcation diagram was obtained in the exact analytical form, which is a
rare possibility. In other variant of the 1D model, with the Gaussian
profile of the local transverse linear coupling, the results were obtained
by means of the VA (variational approximation). The VA was also used to
solve the bifurcation problem in the system combining the linear and
nonlinear localized transverse coupling. The set of the results reported in
the paper provides for a comprehensive description of the symmetry breaking
in the H-shaped system. In particular, adding the nonlinear coupling to the
1D model makes the characteristics of the symmetry breaking closer to those
found in the full 2D model, which automatically incorporates linear and
nonlinear coupling effects.

The settings considered in this work can be realized in the self-attractive
BEC, and in self-focusing bulk optical media, with the appropriate
transverse profile of the refractive index. It is relevant to stress that
the description of the BEC based on the coupled GPEs is entirely based on
the mean-field approximation. While it is well known that this approximation
provides for an exceptionally accurate description of virtually all
matter-wave patterns observed in experiments, quantum fluctuations and other
beyond-mean-field effects being detectable only under special conditions
\cite{BEC}, one may expect that fluctuations can be amplified in a vicinity
of the phase transition (i.e., \ of the SBB), which suggests a question,
whether the GPEs are applicable in this case. A full analysis of this issue
should be a subject for a separate extended work; nevertheless, a relevant
analogy is suggested by the analysis of fluctuations in the vicinity of the
SBB, both in the continuous-wave and soliton settings, which was performed
in the framework of the model of the directional coupler in nonlinear
optics, based on the coupled 1D NLSEs, in Ref. \cite{Amir}. The conclusion
was that, on the contrary to the anticipations, the fluctuational effects
remain virtually negligible in that case, their amplification in the
vicinity of the respective SBBs producing no conspicuous changes against the
mean-field description. Based on this analogy, we expect that the mean-field
description remains valid in the present case too.

One possibility for further development of the analysis is to perform it in
a two-component system. Another interesting extension may be carried out for
a configuration with \emph{two} parallel transverse rungs, in which case one
may expect a \emph{double} spontaneous symmetry breaking: between the
parallel potential troughs, and, independently, between vicinities of the
two rungs. In particular, a challenging problem would be to find an exact
solution in the case when the pair of the rungs is represented by a
symmetric set of two $\delta $-functions.

\end{document}